\documentclass[fleqn,twoside]{article}
\usepackage{espcrc2}

\usepackage{graphicx}
\usepackage[figuresright]{rotating}

\newcommand{\AmS}{{\protect\the\textfont2
  A\kern-.1667em\lower.5ex\hbox{M}\kern-.125emS}}
\usepackage{bm}
\def\dis{\displaystyle}
\def\itmb{\begin{itemize}}
\def\itme{\end{itemize}}
\def\enmb{\begin{enumerate}}
\def\enme{\end{enumerate}}
\def\eqnb{\begin{equation}}
\def\eqne{\end{equation}}

\def\NPA{{Nucl. Phys.} {\bf A}}
\def\NPB{{Nucl. Phys.} {\bf B}}

\def\PRD{{Phys. Rev.} D}

\title{ Numerical Study of Lattice Landau Gauge QCD \\
and the Gribov Copy Problem}

\author{Hideo Nakajima\thanks{e-mail nakajima@is.utsunomiya-u.ac.jp}\\ 
Department of Information Science, Utsunomiya University, 321-8585 Japan \\
Sadataka Furui \thanks{e-mail furui@umb.teikyo-u.ac.jp}\\
School of Science and Engineering, Teikyo University, 320-8551 Japan}
       
\begin{document}

\begin{abstract}
The infrared properties of lattice Landau gauge QCD of SU(3) are
studied by measuring gluon propagator, ghost propagator, QCD running coupling
and Kugo-Ojima parameter of $\beta=6.0, 16^4,24^4,32^4$ and $\beta=6.4, 
32^4, 48^4, 56^4$ lattices. 
By the larger lattice measurements, we observe that  the runnning coupling measured by the product of the gluon dressing function and the ghost dressing 
function squared rescaled to the perturbative QCD results near the highest
lattice momentum has the maximum of about 2.2 at around $q=0.5$ GeV/c, 
and behaves either approaching constant or even decreasing as $q$ approaches 
zero. The magnitude of the Kugo-Ojima parameter is getting larger but 
staying around $-0.83$ in contrast to the expected value $-1$ 
in the continuum theory. We observe, however, there is an exceptional sample
which has larger magnitude of the Kugo-Ojima parameter and stronger infrared
singularity of the ghost propagator. The reflection positivity of the 1-d Fourier transform of the gluon propagator of the exceptional sample is manifestly violated.

Gribov noise problem was studied by performing the fundamental modular 
gauge (FMG) fixing with use of the parallel tempering method of 
$\beta=2.2, 16^4$ SU(2) configurations. Findings are that the gluon propagator 
almost does not suffer noises, but the Kugo-Ojima parameter and the ghost 
propagator in the FMG becomes $\sim 5$\% less in the infrared region than 
those suffering noises. It is expected that these qualitative aspects 
seen in SU(2) will reflect in the infrared properties of SU(3) QCD as well.

\vspace{1pc}
\end{abstract}

\maketitle

\section{Introduction}

One of our basic motivations in the present study is verification of
the color confinement mechanism in the Landau gauge.
Two decades ago, Kugo and Ojima proposed a criterion for the color confinement
in Landau gauge QCD using the Becchi-Rouet-Stora-Tyutin(BRST) invariance of  continuum theory \cite{KO}. Gribov pointed out that the Landau gauge can not
be uniquely fixed, that is, the Gribov copy problem, and argued that the 
unique choice of the gauge copy could be a cause of the color confinement 
\cite{Gr}. Later Zwanziger developed extensively the lattice 
Landau gauge formulation \cite{Zw} in view of the Gribov copy problem. 
Kugo and Ojima started from naive Faddeev-Popov Lagrangian obviously
ignoring the Gribov copy problem, and gave the color confinement criterion
with use of the following two point function of the
covariant derivative of the ghost and the commutator of the antighost 
and gauge field: 
\begin{eqnarray}
(\delta_{\mu\nu}-{q_\mu q_\nu\over q^2})u^{ab}(q^2)\qquad\qquad\qquad\qquad\qquad\qquad
&&\nonumber\\
={1\over V}
\sum_{x,y} e^{-ip(x-y)}\langle {\rm tr}\left({\lambda^a}^{\dag}
D_\mu \displaystyle{1\over -\partial D}[A_\nu,\lambda^b] \right)_{xy}\rangle&&,
\label{KOCR}
\end{eqnarray}
where lattice simulation counterpart is utilized. They claim that
sufficient condition of the color confinement is that $u(0)=-1$ with 
$u^{ab}(0)=\delta^{ab} u(0)$. 
Kugo showed that
\begin{equation}
1+u(0)=\frac{Z_1}{Z_3}=\frac{\tilde Z_1}{\tilde Z_3},
\label{EQSLV}
\end{equation}
where $Z_3$ is the gluon wave function renormalization factor, $Z_1$ is the gluon vertex renormalization factor, and $\tilde Z_3$ is the ghost wave function renormalization factor, respectively.  In the continuum theory $\tilde Z_1$ is
a constant in perturbation theory and is set to be 1. On the lattice, it is
not evident that it remains 1 when strong non-perturbative effects are present.  In a recent SU(2) lattice simulation with several values of $\beta$, finiteness of $\tilde Z_1$ seems to be confirmed, but its value differ from 1. 


The same equality, the first one of 
equations (\ref{EQSLV}), was derived by Zwanziger with use of his "horizon condition" about the same time \cite{Zw}. It is to be noted that 
arguments of both Kugo and Zwanziger are perturbative ones in that they used diagramatic expansion, and the equation (\ref{EQSLV}) is of continuum theory
or continuum limit.

The non-perturbative color confinement mechanism was studied with 
the Dyson-Schwinger approach \cite{SHA,Blo1} and lattice simulations 
\cite{adelaide,orsay1,NF,lat03,FN03}. 
Both types of studies are complementary in that Dyson-Schwinger approach needs
ansatz for truncation of interaction kernels and lattice simulation is
hard to draw conclusions of continuum limit although the calculation is
one from the first principle.

We measured in SU(3) lattice Landau gauge simulation with use of
two options of gauge field definition ($\log U$, $U$ linear; see below), 
gluon propagator, ghost propagator, QCD running coupling
and Kugo-Ojima parameter of $\beta=6.0, 16^4,24^4,32^4$ and $\beta=6.4, 
32^4, 48^4, 56^4$ lattices. The QCD running coupling $\alpha_s=g^2/4\pi$ can be measured in terms of gluon dressing functiuon $Z(q^2)$ and ghost dressing function $G(q^2)$, as renormalization group invariant quatity $g^2 G(q^2)^2 Z(q^2)$.
Infrared features of $g^2$ is not known, however, and there remains a problem of checking the Gribov noise effect among those
quantities, since there exist no practical algorithms available 
so far for fixing fundamaental modular gauge (see below). In our $56^4$ 
simulation, we encountered a copy of an exceptional configuration yielding 
extraordinarily large Kugo-Ojima marameter $c=-u(0)$, and studied 
its feature in some more detail. We made another copy by adjusting 
controlling parameter in gauge fixing algorithm, and measured 
copywise 1-d FT of the gluon propagator, and found violation of reflection
positivity in both cases.


In order to study the Gribov copy problem, we made use of parallel 
tempering \cite{lat03} with 24 replicas to fix fundamental modular 
gauge in SU(2), $\beta=2.2,\ 16^4$ lattice, and 
obtained qualitatively similar result as Cucchieri\cite{Cch}.

\subsection {The lattice Landau gauge}

We adopt two types of the gauge field definitions:
\enmb
\item $\log U$ type:
$U_{x,\mu}=e^{A_{x,\mu}},\ A_{x,\mu}^{\dag}=-A_{x,\mu},$
\item $U$ linear type:
$A_{x,\mu}=\displaystyle{1\over 2}(U_{x,\mu}-U_{x,\mu}^{\dag})|_{trlp.}$,
\enme
where $_{trlp.}$ implies traceless part.
The Landau gauge, $\partial A^g=0$ is specified as a stationary point 
of some optimizing functions $F_U(g)$ along gauge orbit where $g$ denotes 
gauge transformation, i.e., $\delta F_U(g)=0\ {\rm for\ any\ } \delta g.$

Here $F_U(g)$ for the two options are \cite{FN03,Zw}
\enmb
\item 
$F_U(g)=||A^g||^2=\sum_{x,\mu}{\rm tr}
 \left({{A^g}_{x,\mu}}^{\dag}A^g_{x,\mu}\right)$,
\item
$F_U(g)=\sum_{x,\mu}\left (1- {1\over 3}{\rm Re}\ {\rm tr}U^g_{x,\mu}\right),$
\enme
respectively. Under infinitesimal gauge transformation
$g^{-1}\delta g=\epsilon$, its variation reads for either defintion as 
\[
\Delta F_U(g)=-2\langle \partial A^g|\epsilon\rangle+
\langle \epsilon|-\partial { D(U^g)}|\epsilon\rangle+\cdots,
\]
where the covariant derivativative $D_{\mu}(U)$ for two options reads 
commonly as\\
$
D_{\mu}(U_{x,\mu})\phi=S(U_{x,\mu})\partial_\mu \phi+[A_{x,\mu},\bar \phi]
$\\
where 
$
\partial_\mu \phi=\phi(x+\mu)-\phi(x)$, and\\
$\bar \phi=\dis{\phi(x+\mu)+\phi(x)\over 2}
$, but definition of operation $S(U_{x,\mu})B_{x,\mu}$ is given
\enmb
\item
\eqnb
S(U_{x,\mu})B_{x,\mu}=T({\cal A}_{x,\mu})B_{x,\mu}
\eqne
where ${\cal A}_{x,\mu}=adj_{A_{x,\mu}}=[A_{x,\mu},\cdot]$,\\ 
$
T(x)=\dis{{x/2\over {\rm th}(x/2)}}.
$
\item
\eqnb
S(U_{x,\mu})B_{x,\mu}=\dis{1\over 2}\left.\left\{ \dis{U_{x,\mu}+U_{x,\mu}^\dag\over 2},B_{x,\mu}\right\}\right |_{trlp.}
\eqne
\enme

The {\bf fundamental modular gauge} (FMG) \cite{Zw} is specified by the
{\bf global minimum} of the minimizing function $F_U(g)$ along the gauge 
orbits in either case, i.e., \\
$\Lambda=\{U|\ A=A(U), F_{U}(1)={\rm Min}_gF_{U}(g)\}$, 
$\Lambda\subset \Omega$, 
where $\Omega$ is {\bf Gribov region} (local minima), and 
$\Omega=\{U|-\partial { D(U)}\ge 0\ ,\ \partial A=0\}.$

\section{Numerical simulation of lattice Landau gauge QCD}

\subsection{Method of simulation}
First we produce Monte Carlo (Boltzmann) samples of link variable 
configuration
according to Wilson's plaquette action by using the heat-bath method.
We use optimally tuned combined algorithm of 
Creutz's and Kennedy-Pendleton's for SU(2) case, and 
Cabbibo -Marinari pseudo-heat-bath method  with use of 
the above SU(2) algorithm for SU(3) case. As for FMG fixing, the
smearing gauge fixing works well for large $\beta$ on 
relatively small size lattice \cite{NF}. However we found that
its performance for SU(3), $\beta=6.0$, $16^4$ lattice is not
perfect in comparison with our standard method for $\log U$ type definition, 
in which the  third order perturbative treatment of the linear equation with respect to gauge field $A$, 
$-\partial D \epsilon=\partial A$ is performed.
For $U$ linear definition, we use the standard overrelaxation method for
both SU(2) and SU(3). Thus we know that our gauge fixing is not FMG fixng.
The  accuracy of $\partial A(U)=0$ is $10^{-4}$ in the maximum norm in 
all cases.

Measurement of quantity in question, gluon propagator, ghost propagator,
Kugo-Ojima parameter, etc., is performed with use of Landau gauge fixed copy,
and averaged over Monte Carlo samples.

We define the gluon dressing function $Z_A(q^2)$ from the gluon propagator of SU($n$) 
\begin{eqnarray}
D_{\mu\nu}(q)&=&{1\over n^2-1}\sum_{x={\bf x},t}e^{-ikx}{\rm tr}
\langle A_\mu(x)^\dagger A_\nu(0)\rangle \nonumber\\
&=&(\delta_{\mu\nu}-{q_\mu q_\nu\over q^2})D_A(q^2),
\label{GLP}
\end{eqnarray}
as $Z_A(q^2)=q^2 D_A(q^2)$.

The ghost propagator is defined as the Fourier transform of an expectation value of the inverse of Faddeev-Popov(FP) operator ${\cal  M}=-\partial D$,
\begin{equation}
D_G^{ab}(x,y)=\langle {\rm tr} \langle {\lambda^a}_x|({\cal  M}[U])^{-1}|
{\lambda^b}_y\rangle \rangle,
\label{GHP}
\end{equation}
where the outmost $\langle\cdots \rangle$ denotes average over samples $U$. 

\subsection{Numerical results of simulation}
$D_A(q^2)$ is defined in (\ref{GLP}), and its dressing function is given as 
$Z_A(q^2)=q^2 D_A(q^2)$. 
\begin{figure}[htb]
\begin{center}
\includegraphics{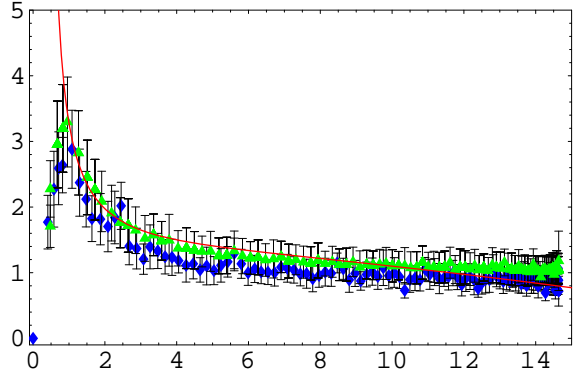}
\end{center}
\caption{The gluon dressing function as the function of the momentum $q$(GeV).  $\beta=6.4$, $48^4$(stars) and $56^4$(diamonds) in the $\log U$ definition. The solid line is that of the $\widetilde{\rm MOM}$ scheme.}\label{gld4856a}
\end{figure}
Numerical results are given in figure.\ref{gld4856a}.

The ghost propagator is defined in (\ref{GHP}). It should be noted that 
the operator $[{\cal M}(U)]^{ab}_{xy}$ is a real symmetric operator 
when $\partial A=0$, and it transforms covariantly
under global color gauge transformation $g$, i.e.,  
$[{\cal M}(U^g)]^{ab}=[g^\dag\{{\cal M}(U)\}g]^{ab}$. 
The conjugate gradient method is used for the evaluation the ghost
propagator, and absence of color off-diagonal components manifests itself,
a signal of no color summetry violation in Landau gauge.
The accuracy of the inversion is maintained less than 5\% in maximum norm
in error check of the source term. \\

We measured the running coupling from the product of the gluon dressing 
function and the ghost dressing function squared.
\begin{equation}
\alpha_s(q^2)=\frac{g_0^2}{4\pi}Z(q^2){ G(q^2)}^2\simeq (qa)^{-2(\alpha_D+2\alpha_G)}.
\end{equation}

\begin{figure}[htb]
\begin{center}
\includegraphics{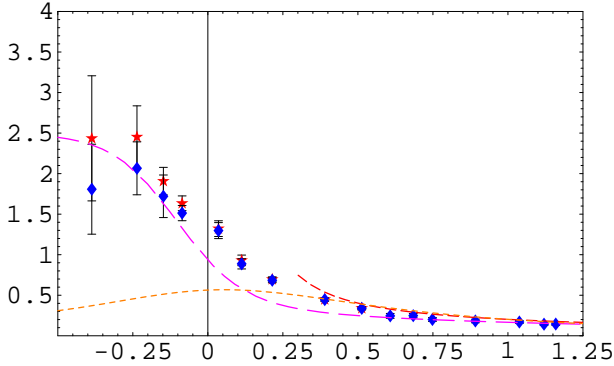}
\end{center}
\caption{The running coupling $\alpha_s(q)$ as a function of the logarithm of momentum $\log_{10}q$(GeV) of the $\log U$, $\beta=6.4$, $56^4$ lattice using the ghost propagator of the average (diamond). The result using the ghost propagator of $I_A$ copy (star) is also plotted for comparison. The DSE approach (long dashed line), the perturbative QCD+$c/q^2$ (short dashed line) and the contour improved perturbation method (dotted line) are also shown.
}\label{alp56}
\end{figure}
The lattice data (diamonds) are compared with a fit of the DSE approach\cite{Blo1} with infrared fixed point $\alpha_0=2.5$, and the pertubative QCD with $c/q^2$ correction \cite{orsay1} and the contour improved perturbation method\cite{FN03}. The The result using the ghost propagator of $I_A$ copy (star) is also plotted for presenting the ambiguity due to the Gribov copy.

Kugo-Ojima color confinement criterion is given with use of $u^{ab}(p^2)$ 
defined in (\ref{KOCR}) with $
u^{ab}(p^2)=\delta^{ab}u(p^2)
$, as $u(0)=-c;\ c=1 \to {\rm color\  confinement}$.
Zwanziger's horizon condition \cite{Zw} derived in the
infinite volume limit is written as follows;
\[
\sum_{x,y} e^{-ip(x-y)} \left \langle {\rm tr}\left({\lambda^a}^{\dag}
D_\mu \displaystyle{1\over -\partial D}(-D_\nu)\lambda^b\right)_{xy}\right
\rangle
\]
\[
=G_{\mu\nu}(p)\delta^{ab}
=\left(\displaystyle{e\over d}\right)\displaystyle{p_\mu p_\nu\over p^2}\delta^{ab}
-\left(\delta_{\mu\nu}-\displaystyle{p_\mu p_\nu\over p^2}
\right)u^{ab},
\]
where $e=\left\langle\sum_{x,\mu}{\rm tr}(\lambda^{a\dag} 
S(U_{x,\mu})\lambda^a)\right\rangle/\{(n^2-1)V\}$,
and the horizon condition reads $\displaystyle \lim_{p\to 0}G_{\mu\mu}(p)-e=0$,
and the l.h.s. of the condition is 
$
\left(\displaystyle{e\over d}\right)+(d-1)c-e=(d-1)h
$
where $h=c-\dis{e\over d}$ and dimension $d=4$, and it follows that 
$h=0 \to {\rm horizon\  condition}$, and thus the horizon condition coincides
with Kugo-Ojima criterion provided the covariant derivative approaches
the naive continuum limit, i.e., $e/d=1$.\\
Values of 
$\displaystyle c, \frac{e}{d}, h$ are shown in the Table \ref{kgtab}.(suffix 1: $\log U$, suffix 2: $U-$linear)

\begin{table*}[htb]
\caption{The Kugo-Ojima parameter $c$ in $\log U$ and $U-$linear version. $\beta=6.0$ and $6.4$. }\label{kgtab}
\begin{tabular}{c|c|ccc|ccc}
 $\beta$&$L$ &$c_1$ & $e_1/d$ & $h_1$ & $c_2$ & $e_2/d$ & $h_2$ \\
\hline
6.0 &16 & 0.628(94)& 0.943(1) & -0.32(9)&  0.576(79) &   0.860(1) & -0.28(8)\\
6.0 &24 & 0.774(76)& 0.944(1) & -0.17(8)&  0.695(63)  &  0.861(1) & -0.17(6)\\
6.0 &32 & 0.777(46)& 0.944(1) & -0.16(5)&  0.706(39)  &  0.862(1) & -0.15(4)\\
\hline
6.4 &32 & 0.700(42)& 0.953(1) & -0.25(4)& 0.650(39) & 0.883(1) & -0.23(4)\\
6.4 &48 & 0.793(61)& 0.954(1) & -0.16(6)& 0.739(65) & 0.884(1) & -0.15(7)\\
6.4 &56 & 0.827(27)& 0.954(1) & -0.12(3)& 0.758(52) & 0.884(1) & -0.13(5)\\
\end{tabular}
\end{table*}

Among gauge fixed configurations, $\beta=6.4$, $\log U$ type, there appeared
a configuration ($I_A$) which have exceptionally large Kugo-Ojima parameters, 
$c=0.851(77)$ with rather large fluctuation among diagonal components. 
For more informtion, we made another copy of it by changing some parameter
of the gauge fixing program. They have very close values in $\|A\|^2$ norm,
but fairly large difference in physically important values. One dimensional Fourier transforms transverse to 4 axes were calculated with respect to sample-wise transverse gluon propagator functions of momentum $q$. Existence of negative value region in one dimensional Fourier 
tranform of (sample averaged) propagator implies violation of reflection 
positivity postulate\cite{FN04}. 

\section{The Gribov copy problem}
The Gribov copy problem is one of fundamental problems in QCD.
FMG is the unique gauge known as the legitimate Landau gauge without Gribov copies. No algorithms for FMG fixing have been established.
Numerical invesigations of Gribov noise have been done in a few 
works \cite{Cch}. For $U$ linear type gauge field definition, 
FMG gauge fixing with use of parallel tempering (PT)
with 24 replicas was performed for $\beta=2.2, 16^4$, SU(2) 67 Monte Carlo configurations. Some more details of the method are given in \cite{lat03}.
We compared data between two groups, the absolute minimum configurations 
of $F_U(g)$ obtained by Landau gauge fixing via PT and 
the 1st copies obtained by the direct application of 
Landau gauge fixing on Monte Carlo configurations. 
The results are as follows; 1)The singularity of the ghost propagator of the FMG
is about 6\% less than that of the 1st copy. 2)Kugo-Ojima parameter $c$ of the
FMG is about 4\% smaller than that of the 1st copy.
3) The gluon propagator of the two groups are almost the same within 
statistical errors. 4) The running coupling of FMG is about 10\% 
suppressed than that of the 1st copy.
5) The horizon function deviation parameter $h$ of the 
FMG is not closer to 0, i.e. the value expected in the continuum limit, 
than that of the 1st copy. 

In Fig.\ref{f-kg8n}, shown is the scatter plot of Gribov copies on the plane of 
Kugo-Ojima parameter vs $1-F_U(g)$, 32 copies via random gauge transformation 
and one copy given via PT, all being Gribov copies on a single gauge orbit.\\
\begin{center}
\begin{figure}[htb]
\includegraphics{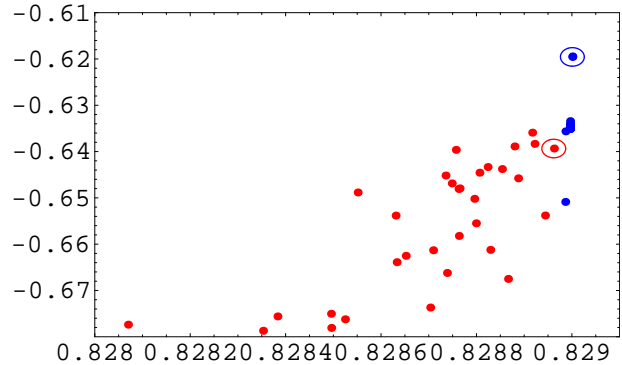}
\caption{The Kugo-Ojima parameter $u(0)$ vs $1-F_U(g)$ of the 33 copies on a single gauge orbit. 
 $SU(2)$,@$\beta=2.2$, $16^4$. The maximum optimization point by Cucchieri's
prescription(lower circle) and our prescription(upper circle) are indicated.}\label{f-kg8n}
\end{figure}
\begin{figure}[htb]
\includegraphics{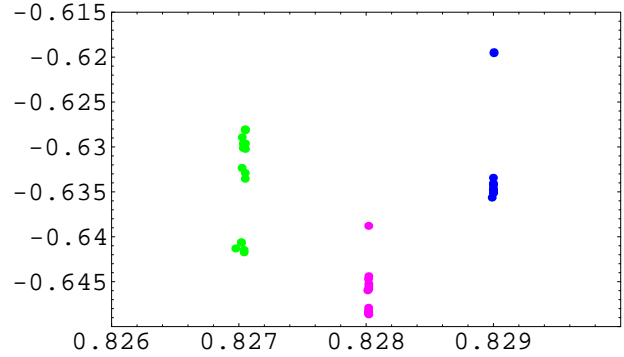}
\caption{The Kugo-Ojima parameter $u(0)$ vs $1-F_U(g)$ of 16 Gribov copies via PT for three gauge orbits each. }\label{f-kg}
\end{figure}
\end{center}

In Fig.\ref{f-kg}, the best one of 16 copies should be picked up for the FMG.
This figure shows how flat is the bottom of $F_U(g)$ with respect to the 
physical quantity.

\section{Discussion and Conclusion} 
The SU(3) $\beta=6.0, 16^4, 24^4, 32^4$ and $\beta=6.4, 32^4, 48^4, 56^4$ lattice Landau gauge simulation was performed. A comparison of the data of $\log U$ type and the $U-$linear type reveals that the gluon propagator does not depend on the type and the ghost propagator of the $U-$linear is about 14\% larger than the $\log U$. The Kugo-Ojima parameter $c$ is still getting larger $u(0)\sim -0.83$ in $\log U$ and about -0.76 in $U-$linear, in accordance to lattice size. 
The running coupling $\alpha_s(q)$ of the $\log U$ type has a maximum of about 2.2, and that of the $U-$linear is slightly smaller since they are rescaled at the high lattice momentum region.
The absolute value of the index $\alpha_G$ increases as the lattice size becomes large but since it is defined at $q\sim 0.4GeV$ region, it is smaller than $\kappa\sim 0.5$ of the DSE which is defined at the 0 momentum region by about factor 2. 

As is studied in SU(2) FMG data, the gluon propagator suffers almost 
no Gribov noise, but magnitude of Kugo-Ojima parameter becomes smaller in FMG, 
and the ghost propagator exhibits less singular infrared behavior in FMG than 
that supposed suffering noise. Accordingly the running coupling becomes smaller
in FMG. All samples in SU(3) can not be supposed in FMG and supposed to 
suffer the Gribov noise, and it is expected that these qualitative aspects 
seen in SU(2) will reflect in the infrared behavior of SU(3) QCD as well. 
The FMG is mathematically well defined on lattice, and its existence
can be proven. But rather flat valued feature of the FMG optimizing function 
with respect to physical quantities seem to keep annoying us, at least, technically, and this fact may suggest necessity of some new formulation of
infrared dynamics of QCD.
In the Langevin formulation of Landau gauge QCD, Zwanziger conjectured 
that the path integral over the FM region will become equivalent to that 
over the Gribov region in the continuum \cite{Zw1}. If his conjecture is
correct, then in order for the gauge dependent quantity, e.g., 
the optimizing function to have the same expectation values on both regions, 
the situation that the probability density may be concentrated in 
some common local region becomes favorable one. The proximity of the FM 
region and the boundary of the Gribov region in SU(2)
in $8^4, 12^4$ and $16^4$ lattices with $\beta=0, 0.8, 1.6$ and $2.7$ was 
studied  in \cite{Cucc}. The tendency that the smallest eigenvalue of the 
Faddeev-Popov matrix of the FMG and that of the 1st copy come closer as 
$\beta$ and lattice size become larger was observed, although as remarked 
in \cite{Cucc} the physical volume of $\beta=2.7$, $16^4$ lattice is small 
and not close to the continuum limit. In this respect, further study of 
lattice size and $\beta$ dependence of Gribov noise of various quantities will
become an important problem in future. 

Our observation of exceptional 
configurations in SU(3), $\beta=6.4$, $56^4$ may be considered as 
a signal of the tendency of the probabililty concentration to the 
Gribov boundary\cite{Zw,FN04}.

The confinement scenario was recently reviewed in the framework of the renormalization 
group equation and dispersion relation \cite{Kond}. It is shown that the gluon dressing function
satisfies the superconvergence relation, and the gluon propagator does not necessarily vanish as Gribov and Zwanziger conjectured, but it should be finite. 
The multiplicative renormalizable DSE approach\cite{Blo1} predicts that the exponent $\kappa=0.5$ and the infrared fixed point $\alpha_0=2.6$ and our lattice data are consistent with this prediction.

{\bf Acknowledgments:}

We are grateful to Daniel Zwanziger for enlightning discussion.
This work is supported by the KEK supercomputing project No. 03-94,
and No. 04-106.


\begin{thebibliography}{99}
\bibitem{KO} T. Kugo and I. Ojima, {Prog. Theor. Phys. Suppl.} {\bf 66}, 1 (1979).

\bibitem{Gr} V.N. Gribov, {\NPB} {\bf 139}{1}{(1978)}.

\bibitem{Zw} D. Zwanziger, {\NPB} {\bf 364} ,{127} {(1991)}, idem B
{\bf 412}, {657} (1994).


\bibitem{SHA} L. von Smekal, A. Hauck,  R. Alkofer, {Ann. Phys.} (N.Y.) {\bf 267},1 (1998).

\bibitem{Blo1} J.C.R. Bloch,  Few Body Syst {\bf 33},{111}{(2003)}.

\bibitem{adelaide} D.B. Leinweber et al.,  {\PRD}{\bf 60},{094507}{(1999)}; ibid {\PRD}{\bf 61},{079901}{(2000)}.

\bibitem{orsay1} D. Becirevic et al., {\PRD} {\bf 61},{114508}{(2000)}.
\bibitem{NF} H.Nakajima and S. Furui, {\NPB} (Proc Suppl.){\bf 73A-C},{635, 865}(1999) and references therein, 
 {\NPB} (Proc Suppl.){\bf 83-84},521 (2000), {\bf 119},730(2003);  
{\NPA} {\bf 680},{151c}(2000).



\bibitem{lat03} H. Nakajima and S. Furui, {\NPB}(Proc. Supl.)
{\bf 129-130}(2004) and references therein, hep-lat/0309165.
\bibitem{FN03} S. Furui and H.Nakajima, {\PRD}{\bf 69},{074505}{(2004)} and references therein.
\bibitem{FN04} S. Furui and H.Nakajima, hep-lat/0403021 ver 3.


\bibitem{Cch} A.Cucchieri, {\NPB}{\bf 508},353{(1997)},\\ hep-lat/9705005.
\bibitem{Zw1} D. Zwanziger, {\PRD}{\bf 69},{016002}{(2004)}, hep-ph/0303028.

\bibitem{Cucc} A. Cucchieri, {\NPB}{\bf 521},{365}{(1998)}, hep-lat/9711024.

\bibitem{Kond} K.I. Kondo, hep-th/0303251.
\end{thebibliography}
\end{document}